\pdfoutput=1
\documentclass[aps,prl,twocolumn,nofootinbib,showpacs,superscriptaddress,groupedaddresst,preprintnumbers,12pt]{revtex4-1}
\usepackage{graphics,graphicx,psfrag}
\usepackage{amsmath,amssymb}
\usepackage[sort&compress]{natbib}
\usepackage{bm}	
\usepackage{verbatim}
\usepackage{multirow}
\usepackage{subfigure}
\usepackage{ulem}
\usepackage{color}
\usepackage[linktocpage=true,bookmarksnumbered=true,colorlinks=true,plainpages,linkcolor=blue,citecolor=red]{hyperref}
\usepackage[utf8]{inputenc}  

\def\Zp{Z^{\prime}}

\def\issue(#1,#2,#3){{\bf #1}, #2 (#3)}

\catcode`\@=11
\def\lsim{\mathrel{\mathpalette\@versim<}}
\def\gsim{\mathrel{\mathpalette\@versim>}}
\def\@versim#1#2{\vcenter{\offinterlineskip
\ialign{$\m@th#1\hfil##\hfil$\crcr#2\crcr\sim\crcr } }}

\catcode`@=12

\newcommand{\newc}{\newcommand}
\newc{\wt}{\widetilde}
\newc{\ra}{\rightarrow}
\usepackage{hyperref}
\def\beq {\begin{equation}}
\def\eeq {\end{equation}}
\def\bi {\begin{itemize}}
\def\ei {\end{itemize}}
\def\bea {\begin{eqnarray}}
\def\eea {\end{eqnarray}}

\newcommand{\br}{\begin{eqnarray}}
\newcommand{\er}{\end{eqnarray}}
\newcommand{\be}{\begin{equation}}
\newcommand{\ee}{\end{equation}}

\graphicspath{{./figs/}}

\usepackage{lineno}
\def\Zp{Z^{\prime}}

\def\issue(#1,#2,#3){{\bf #1}, #2 (#3)}

\hypersetup{
   colorlinks=true,       
   linkcolor=blue,        
   citecolor=red,         
   filecolor=magenta,      
   }

\begin{document}

\preprint{
{\vbox {
\hbox{\bf MSUHEP-17-018}
\hbox{\today}
}}}
\vspace*{2cm}

\title{Characterizing Boosted Dijet Resonances with Jet Energy Correlators }
\vspace*{0.25in}   
\author{R. Sekhar Chivukula}
\affiliation{\vspace*{0.1in}
 Department of Physics and Astronomy
Michigan State University, East Lansing, MI, U.S.A.\\
}
\author{Kirtimaan A. Mohan}
\affiliation{\vspace*{0.1in}
 Department of Physics and Astronomy
Michigan State University, East Lansing, MI, U.S.A.\\
}
\author{Dipan Sengupta}
\affiliation{\vspace*{0.1in}
 Department of Physics and Astronomy
Michigan State University, East Lansing, MI, U.S.A.\\
}
\author{Elizabeth H. Simmons}
\affiliation{\vspace*{0.1in}
 Department of Physics and Astronomy
Michigan State University, East Lansing, MI, U.S.A.\\
}
\affiliation{\vspace*{0.1in} University of California San Diego,  La Jolla, CA, U.S.A.}

\begin{abstract}{
		We show that Jet Energy Correlation variables can be used effectively to discover and distinguish a wide variety of boosted light dijet resonances at the LHC through sensitivity to their transverse momentum and color structures. 
		
}
\end{abstract}
\maketitle

The LHC is actively seeking dijet resonances.  However, for a given resonance mass, the ability to probe smaller couplings to quarks and gluons depends on the amount of data collected and how well one can reduce Standard Model (SM) backgrounds.
Sensitivity to light dijet resonances at the LHC, in particular,  is limited by the presence of large SM backgrounds that accumulate at a rate which is difficult to manage by currently available trigger and data acquisition systems at ATLAS and CMS. 
Looking for such resonances produced with high transverse momenta in association with a jet, photon, $W^{\pm}$ or $Z$ boson (or even in pair production of the resonances) can reduce both signal and  background rates thus avoiding trigger threshold limitations. Additionally, for highly boosted light resonances, jet substructure techniques can be applied to further reduce backgrounds.

Recently, using this search strategy, ATLAS~\cite{ATLAS-CONF-2016-070} and CMS~\cite{Sirunyan:2017dnz} were able to set limits on narrow light vector resonances (specifically a leptophobic $\Zp$~\cite{Dobrescu:2013coa}), decaying to a pair of jets, in a coupling and mass range $(100-600)~\text{GeV}$ that was not accessible to earlier colliders such as UA2 and CDF.  However there are a plethora of possible dijet resonances that could exist: colorons~\cite{Chivukula:1996yr}, sextet and triplet diquarks~\cite{Hewett:1988xc,Pati:1974yy}, excited quarks~\cite{Cabibbo:1983bk,Baur:1987ga}, color-octet scalars~\cite{Hill:2002ap}, massive spin-2 particles~\cite{Han:1998sg} to name a few.
While substructure techniques can unearth new resonances, once a light resonance is discovered the primary task becomes understanding the nature of the resonance itself.
In this note we demonstrate how Jet Energy Corelators (JECs) aid in differentiating between these numerous types of resonances~\footnote{Elsewhere we will consider and compare other jet observables such as N-Subjetiness~\cite{Thaler:2010tr,Butterworth:2008iy} and Jet Energy Profiles~\cite{Li:2011hy}, jet pull~\cite{Gallicchio:2010sw}  and dipolarity~\cite{Hook:2011cq}.}.
 
 New dijet resonances may be classified according to their spin and color structure \cite{Han:2010rf}. While resonances of different spin can be differentiated on the basis of angular distributions of their decay products, identifying types of resonances on the basis of their color structure is more difficult. Note that the color discriminant variable~\cite{Chivukula:2014npa,Chivukula:2015zma} will not be useful for (relatively) weakly-coupled light resonances since their decay widths are too narrow to be measured at the LHC.
 We exhibit the power of JEC in this regard by examining some benchmark models\footnote{We restrict ourselves to leptophobic models only.} listed below: 
 \begin{itemize}
 	\item A color singlet leptophobic $\Zp$ that couples to baryon number via $\frac{g_B}{6}\bar{q}\gamma^{\mu}q Z^{\prime}_{\mu}$~\cite{Dobrescu:2013coa}.
 	\item A color octet coloron $C_\mu$ interacting with quarks through $g_s \tan{\theta}\ \bar{q}\  T^{a}\gamma^{\mu}q C^{a}_{\mu}$\cite{Chivukula:1996yr}.
 	\item A color sextet diquark $(\Phi^\gamma_6)$ that interacts with pairs of quarks through $\sqrt{2} (\bar{K}_6)^{ab}_\gamma \lambda_\Phi \Phi^\gamma_6 \bar{u}^c_{Ra}u_{Lb} $\cite{Hewett:1988xc,Chivukula:2015zma}. 
 	\item A color triplet  excited quark ($q^*$) interacting with quarks and gluons (as well as other gauge bosons) through the interaction term
 	$\frac{1}{2\Lambda} \bar{q}^{*}_R \sigma^{\mu\nu} [ g_S f_S \frac{\lambda^{a}}{2} G^{a}_{\mu\nu}]q_{L}$  \cite{Baur:1987ga}.
 	\item A massive color singlet, spin 2 object ($X^{\mu\nu}$) that interacts with SM particles through the energy momentum tensor $T_{\mu\nu}$ as $\frac{1}{\Lambda}X^{\mu\nu}T_{\mu\nu}$ \cite{Han:1998sg}.
 	\item A color octet (but electroweak singlet) scalar $S_{8}$ that interacts with gluons through the field strength tensor as  $\frac{g_s d_{ABC} k_s}{\Lambda} S_{8}^A G_{\mu\nu}^{B}G^{C,\mu\nu}$
 	\cite{Hill:2002ap,Dobrescu:2007yp}.
 \end{itemize}
While this list is not exhaustive, these examples serve to illustrate the utility of this method. 
 
 The signal process of interest is the production of various resonances in association with a jet, viz. $(pp \to X(\to jj)+ j\ ; X\epsilon\{\Zp_{\mu},C_{\mu},\Phi_{6},q^*,X^{\mu\nu},S_{8}\})$\footnote{We also performed an analogous analysis of the production of the resonances in association with a W boson, which will be reported in a future work.}, where the resonance is boosted sufficiently that its decay products lie within a single ``fat jet".  The dominant background originates from QCD multijet events.  
The various resonance models were implemented in {\tt Feynrules}~\cite{Alloul:2013bka}.  Parton level events for both signal and background were simulated using  {\tt MADGRAPH\_AMC@NLO}~\cite{Alwall:2014hca} assuming 13 TeV LHC energy, with subsequent showering and hadronization performed using {\tt PYTHIA8}~\cite{Sjostrand:2014zea}.  We use {\tt FASTJET}~\cite{Cacciari:2011ma} to reconstruct jets and calculate JECs. 
Additionally jet energy smearing and detector granularity are simulated using {\tt Delphes3}~\cite{deFavereau:2013fsa} with parameters similar to ATLAS.
We use the Cambridge-Aachen algorithm~\cite{Dokshitzer:1997in} to construct fat-jets of radius $R=1.0$ and use the mass-drop  tagger~\cite{Butterworth:2008iy} to resolve the fat jets into subjets to reconstruct the mass of the resonance $X$ within $\rm M_{X}\pm 20~GeV$ to help reduce the background. 
Importantly, we find that the mass-drop tagger does not significantly affect JEC distributions of unfiltered signal fat-jets. Further, the acceptance of the tagger does not depend significantly on the nature of the resonance. 
We require $\rm H_{T} = \Sigma p_{T} > 900\ \text{GeV}$ and  $p^{\text{fatjet}}_{T} > 500~\text{GeV}$. We use {\tt MCFM}~\cite{Campbell:2015qma} to determine K-factors for NLO production of the $V+$jets, $t\bar{t}$ and single top backgrounds.
NLO K-factors for the dijet production cross-section were determined using {\tt POWHEGBOX}~\cite{Nason:2004rx,Frixione:2007vw,Alioli:2010xd}.
Further, we use the MLM~\cite{Mangano:2002ea} matching procedure in {\tt PYTHIA8} for multi-jet events that were generated in {\tt MADGRAPH\_AMC@NLO}. 

For the purpose of demonstration, the mass of the resonance is set to $M_{X} = 250$~GeV. The current 95 $\%$ CL bound on a 250 GeV leptophobic $Z'$  from $35.9 ~\text{fb}^{-1}$ of $13$~TeV data is $g_b \stackrel{<}{\sim} 1.5$ $(g_q\stackrel{<}{\sim} 0.22)$, compared with an expected bound of $g_b\stackrel{<}{\sim} 1.1$~\cite{Sirunyan:2017dnz}. We therefore consider a $Z'$ resonance with $g_b=0.6$, which is still allowed by the data. For this coupling, we find that the cross section after all cuts is $25$~fb. For all other resonances we adjust the value of the couplings such that for all resonances under consideration, the cross-section after cuts is $25$~fb. We find that our total background is $\sim 50$~pb with the dominant contribution coming from QCD multi-jet processes.
 We find that with our cuts we expect $S/\sqrt{B}\sim 1.9\sigma$ which is comparable to the expectations of experimental results.


%
 
JECs were originally introduced in \cite{Banfi:2004yd,Jankowiak:2011qa} as a two point correlator, and generalized in \cite{Larkoski:2013eya}. Studies on JEC have focused on standard model processes, specially to distinguish quark jets from gluon jets. Additionally, JECs have been shown to be able to differentiate boosted Higgs and top quarks from QCD backgrounds \cite{Larkoski:2013eya}.  
   
 The N-point generalized JEC is defined as \cite{Larkoski:2013eya},
 {\footnotesize
\begin{equation}\label{eq:JEC}
ECF(N,\beta)= \sum_{(i_1<..<i_N\in J)}\prod_{a=1}^{N} p_{T{i_a}}\Bigg(\prod_{b=1}^{N-1}\prod_{c=b+1}^{N}R_{i_{b}i_{c}}\Bigg)^{\beta}.
\end{equation}
}
 The sum runs over all objects (tracks\footnote{Here we define the JECs in terms of the individual particles in the ``fat jet" in the simulated event, after using the detector simulation as noted above.}  or calorimeter cells) within a system J (individual jets or all final states of the collision).  
$p_{T_{i}}$ is the transverse momentum of each constituent object.
The variable $R_{ij}= \sqrt{(\eta_{i}-\eta_{j})^{2} + (\theta_{i} - \theta_{j})^{2}}$, denotes a pairwise distance measure and is raised to the power $\beta$. 
Here $\eta_i$ is the pseudo-rapidity while $\theta_{i}$ is the azimuthal angle of particle $i$.  
The entire function is infrared and collinear safe for  $\beta > 0$. 

\begin{figure}[t]
	\centering
	\includegraphics[width= 0.45 \textwidth]{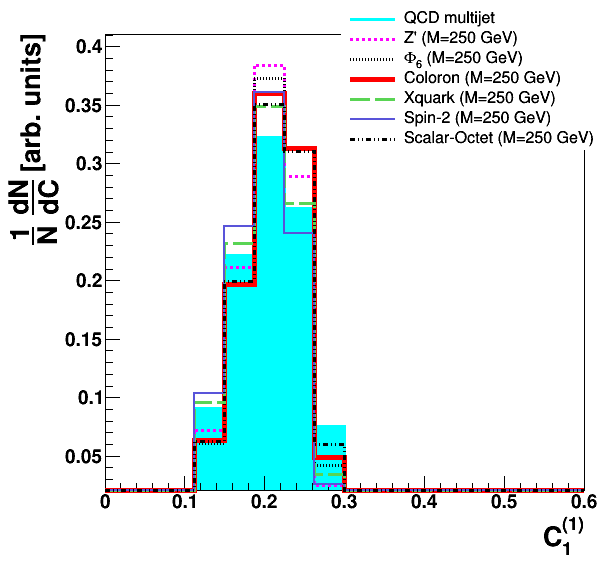}
	\caption{The double ratio distribution for $C_{1}^{(1)}$ for the different kinds of resonances under consideration; $\Zp$ in pink (small-dashed), sextet-diquark $\Phi_{6}$ in black (dotted) , Coloron ($C_{\mu}$) in red (bold,thick), excited quark ($q^{\star}$, Xquark) in green (large-dashed), Spin-2 ($X^{\mu\nu}$) in blue (bold,thin), scalar color octet ($S_8$) in black (dot-dashed) . The cyan shaded region corresponds to the distribution of the multi-jet background.
		\label{fig:c11}}
\end{figure}

Using Eq. \ref{eq:JEC} one can construct a dimensionless double ratio as
 \begin{equation}
 \label{eq:CNbeta}
C_{N}^{(\beta)} = \frac{ECF(N+1,\beta)ECF(N-1,\beta)}{ECF(N,\beta)^{2}}\ .
\end{equation}
In general, $C_{N}^{(\beta)}$
quantifies radiation of higher order $\alpha_{s}^{n}$, emerging out of leading order hard sub-jets.
In a boosted $\Zp\to j_{1} j_{2}$ like system, 
if $C_{2}^{(\beta)} < C_{1}^{(\beta)}$, the fat jet has two resolved hard subjets, and higher order substructure is mostly soft or collinear. With subsequent soft emissions 
of the final state, one can assume $p_{T}^{j_1}\simeq p_{T}^{j_2} >> p_{T}^{j_i}$, where $i> 2$. Thus, the leading approximation can be written as,
 \begin{equation}\label{eq:C11}
C_{1}^{(1)}\simeq R_{12}/4\ .
\end{equation}
 Since $R_{12}\simeq  2 m_{\Zp}/p_{T}^{j}$,  $C_{1}^{(1)}$ is directly related to the boost of the resonance. 
 We show the distribution of $C_{1}^{(1)}$ for various resonances in Fig.~\ref{fig:c11},.
 Since we require $R\le 1.0$ and $p_{T}^{\text{fat-jet}}> 500$~GeV, we see that $C_{1}^{(1)}\lesssim 0.25$. Further, since the $p_{T}$ spectrum is almost identical for all resonances under consideration the distribution for $C_{1}^{(1)}$ look the same.
 The $p_{T}$ distribution for $q^{\star}$ and $X^{\mu\nu}$ are slightly harder (and therefore $C_{1}^{(1)}$ is shifted to smaller values) since their interactions are mediated by  dimension 5 operators.
 We would also like to point out here that information about the initial state and therefore the nature of the resonance can be gleaned by comparing the $p_T$ distribution for cases when the resonance is produced in association with other particles such as a $W$-boson.The lower end of the $C_{1}^{(1)}$ distribution is bounded by detector resolution.  This is the minimal separation between 
 subjets that can be resolved, and is  encoded in our implementation of the mass drop tagger.
  

Higher point moments of the JEC depend crucially on the nature of the resonance, in particular, the color structure not only of the resonance but also its decay products -- in particular, since $C_F< C_A$, a color octet will radiate more widely than a color triplet.
This implies that the correlator double ratios $C_{N}^{(\beta)}$ should in general be larger  for a color octet than a color triplet and smallest for a color singlet.

\begin{figure}[t]
	\centering
	\includegraphics[width= 0.45 \textwidth]{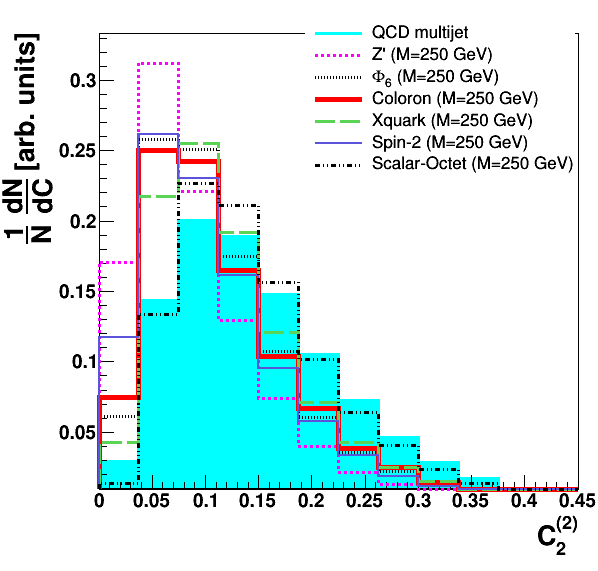}
	\caption{ The double ratio distributions $C_{2}^{(2)}$ for the resonances and the multi-jet background.\label{fig:c22}}
\end{figure}
In Fig.~\ref{fig:c22} we present distributions for the double ratios $C_{2}^{(2)}$. 
To understand the behavior of $C_{2}^{\beta}$,  consider a simplified scenario of the two body hadronic decay of a resonance X
with one soft emission-- 
$\rm X \to 1 + 2 + 3_{\text{soft}}$ where $3_{\text{soft}}$ originates from $1$.
We also expect the distance measure $R_{13}$ to be small and $\rm p_{T}^{j1}\simeq p_{T}^{j2} (=p_{T}) >> p_{T}^{j3}$ in the soft and collinear approximation .
 $C_{2}^{(\beta)}$  can then  be approximated as 
\begin{equation}
C_{2}^{(\beta)}  \simeq \frac{ 2 \varepsilon R_{12}^{\beta}R_{13}^{\beta}R_{23}^{\beta}  } { (R_{12}^{\beta} + \varepsilon R_{13}^{\beta} + \varepsilon R_{23}^{\beta} )^{2} }~;
\end{equation}
note that $\varepsilon R_{13}= (p_{T}^{j3}/{p_{T}})R_{13} \ll 1$ is doubly suppressed since the third jet, $3_{soft}$, is both low-momentum and colinear with jet 1. We therefore expect $C^{(2)}_{2}$ to peak near $0$ as seen in Fig.~\ref{fig:c22}. As discussed earlier, a small $C_2^{(2)}$ implies that the event is mostly a two prong subjet system. 

In Fig.~\ref{fig:c22} we also see, as expected, that the color singlet $\Zp$ has the smallest values for $C_{2}^{(2)}$ whereas, due to the presence of more radiation, the colored objects have larger values.
Although the spin-2 is a color singlet its distribution is not identical to $\Zp$ and instead has larger values of $C_{2}^{(2)}$. This is because the spin-2 predominantly decays to gluons, which themselves produce broader jets (since $C_F < C_A$), whereas the coloron and $\Zp$ decays to quarks, which produce narrower radiation patterns. As expected, the color octet scalar resonance has the largest values of $C_{2}^{(2)}$ since it is itself an octet which decays to a pair of octets (gluons).
Also shown in Fig.~\ref{fig:c22} is the distribution of $C^{2}_{(2)}$ for the dominant multi-jet background. We see that its distribution is significantly different from most of the signal distributions and therefore the JECs can be used not only to discriminate between different signals but also to discriminate signal from background.\footnote{CMS~\cite{Sirunyan:2017dnz} uses JEC in its search to discriminate between a $\Zp$ and background. The behavior of $C_2^{(2)}$, suggests that in addition to enhancing $S/\sqrt{B}$ we can simultaneously use it to discriminate between resonances ($S_8$ being an exception).} The scalar octet behaves most like the QCD multi-jet background since, at low masses, the
background is mostly gluonic in origin.

\begin{figure}[t]
	\centering
	\includegraphics[width= 0.45 \textwidth]{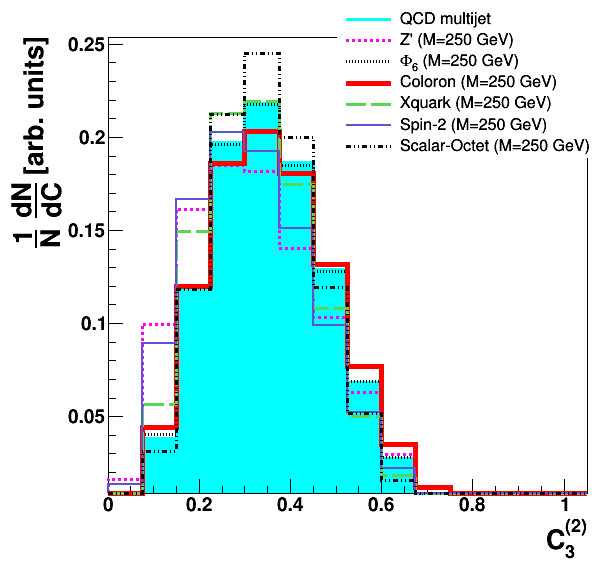}
	\caption{The double ratio distributions for $C_{3}^{(2)}$ for the resonances and the multi-jet background.\label{fig:c23}}
\end{figure}
Further discrimination between resonances can be achieved by looking at the distribution for the higher moment correlator $C_{3}^{(\beta)}$ shown in Fig.~\ref{fig:c23}.
In contrast to $C_{2}^{(\beta))}$ we see that the peak of the distribution is shifted away from $0$.
This behavior can be better understood by considering the scenario where $\rm X \to 1 + 2 + 3_{\text{soft}}  + 4_{\text{soft}}$. 
In this case, we assume that the transverse momentum distribution follows, 
$\rm p_{T}^{j1}\simeq p_{T}^{j2} (=p_{T}) \gg (p_{T}^{j3},   p_{T}^{j4} = p_{T'})$
We can then approximate $C_{3}^{\beta}$ as (up to order $\varepsilon = \frac{p_{T'}}{p_{T}}$) 
\begin{eqnarray}
C_{3}^{(\beta)} & \simeq &  \frac{[(R_{13}R_{14}R_{23}R_{24}R_{34})^{\beta}}{[(R_{13}R_{23} )^{\beta} + (R_{14}R_{24} )^{\beta}]^2} + \mathcal{O}(\epsilon)
\label{eq:C3approx}
\end{eqnarray}
Thus the leading term is not proportional to $\varepsilon$, resulting in the  peak that is shifted away from 0, and is determined by 
the relative opening angles. 
Similar to what we saw for the lower moment correlator, we find that the distribution of $C_{3}^{(2)}$ is shifted to larger values depending on the dimensionality of the $SU(3)$ representation of the resonance as well as its decay products.
The color singlet $\Zp$ decaying to a pair of quarks peaks closer to $0$, whereas  the distribution for the others, which either are octets or decay to gluons, is shifted away from $0$. 
\begin{figure}[t]
	\centering
\includegraphics[width= 0.45 \textwidth]{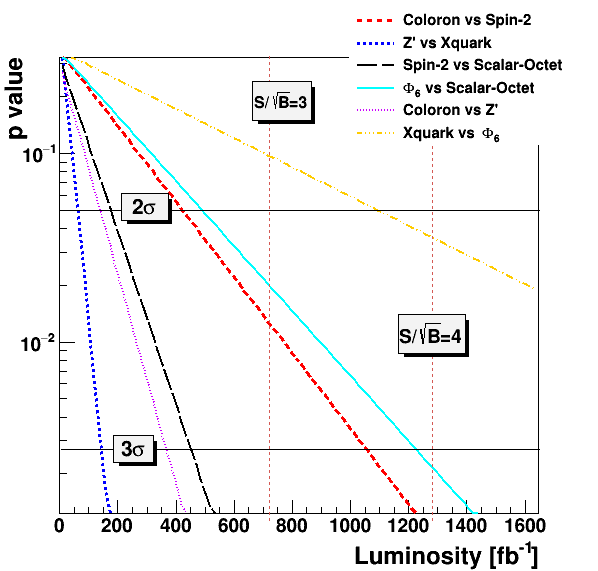}
\caption{ The $p$-values testing hypothetical identities of various resonances as a function of luminosity. Horizontal lines indicate $2$ and $3~\sigma$ exclusion of the alternate hypothesis. Vertical lines show where $S/\sqrt{B} =$ 3 or 4.  }\label{fig:cls}
\end{figure}

An important point that should be noted finally is the dependence of the JEC on the exponent $\beta$. As $\beta \to 0$, the dependence on the relative angles vanishes, and the JEC double ratio approaches an (approximately) constant value away from 0. The exponent should therefore be viewed as a weighting factor that controls the size of the variation of the JEC. Note that we have not optimized $\beta$ for maximal discrimination in this analysis.
Another aspect that we have not investigated and have reserved for future study is the use of JECs (or other jet observables) on unfiltered subjets to identify quark and gluonic jets. The ability to discern the decay products of these resonances would further enhance our ability to pinpoint the nature of the resonance.

In order to test the ability of JECs to characterize the nature of the resonance we perform a multi-variable likelihood analysis. We do not include $C_{1}^{(1)}$ in our likelihood function, since we are trying to test the information provided by radiation patterns and not kinematics. We therefore include only $C_{2}^{(2)}$ and $C_{2}^{(3)}$ in our likelihood function and test the ability of these two jet observables in differentiating the resonances.
The result of our analysis is shown in Fig.~\ref{fig:cls}. The horizontal dotted lines indicate where one can distinguish between various signal hypotheses at the $2~\sigma$ or $3~\sigma$ level; for example, one could tell a $\Zp$ from an excited quark at the $3~\sigma$ level with about $180~\text{fb}^{-1}$ of integrated luminosity. The vertical lines indicate the value $S/\sqrt{B}$ provided by a given integrated luminosity; for instance, achieving $S/\sqrt{B} = 3$ for our resonances (since we assume the signal size is $25~\text{fb}^{-1}$) would require $720~\text{fb}^{-1}$ of data. The figure shows that it is very easy to tell apart a coloron from a $\Zp$, whereas the weakest discrimination is that between a  spin-2 and a diquark.
 
In summary, we conclude that JECs are a powerful tool to both discover and identify new resonances at the LHC.

\begin{acknowledgements}
This material is based in part upon work supported by the National Science Foundation under Grant No. 1519045. We thank Wade Fisher and Joey Huston for useful discussions.
\end{acknowledgements}

\bibliographystyle{apsrev}
\bibliography{refs} 
\end{document}